# Crystallization via tubing microfluidics permits both in situ and ex situ X-ray diffraction

Charline J.J. Gerard[1], Gilles Ferry[2], Laurent M. Vuillard[2], Jean A. Boutin[2], Leonard M. G. Chavas[3], Tiphaine Huet[3], Nathalie Ferte[1], Romain Grossier[1], Nadine Candoni[1], Stéphane Veesler*[1]

[1]*CINaM - CNRS, Aix-Marseille Université, Campus de Luminy, Case 913, F-13288 Marseille Cedex 09 fax : +334 91418916, email : veesler@cinam.univ-mrs.fr*
[2]*Institut de Recherches SERVIER, 125, chemin de ronde, F-78290 Croissy-sur-Seine.*
[3]*Synchrotron SOLEIL Proxima-I, Gif-sur-Yvette, France*

ABSTRACT.
We used a microfluidic platform to address the problems of obtaining diffraction quality crystals and crystal handling during transfer to the X-ray diffractometer. We optimize crystallization conditions of a pharmaceutical protein and collect X-ray data both in situ and ex situ.

## 1. Introduction

Structural biologists need to solve three-dimensional structures of biological macromolecules via X-ray crystallography. Two decisive and rate-limiting steps are obtaining diffraction-quality crystals and handling crystals during transfer to the diffractometer.

Obtaining diffraction-quality crystals is complex and influenced by many parameters (pH, temperature, types of buffer, salts and crystallization agents). Problems in producing suitable crystals can be tackled in two steps: first, screening for favourable crystallization conditions in the phase diagram and second, optimizing crystal growth by developing a specific kinetic path in the phase diagram. Screening is an expensive task, both in terms of time and raw materials. Moreover, when only small quantities of sample materials are available, a suitable experimental tool is essential. Microfluidic techniques, i.e. the control and manipulation of flows at sub-millimetre scale using miniaturized devices called Lab On Chip (LOC)(van der Woerd *et al.*, 2003) are appropriate for automating, miniaturizing and high-throughput crystallization approaches involving multiple operations such as mixing, analysis, separation(Leng J. & Salmon J.B., 2009). LOCs are applied in both fast screening and optimization stages of protein crystallization studies, via the integration of traditional protocols of protein crystallization(Candoni N. *et al.*, 2012). Furthermore, the microfluidics approach suits the stochastic nature of nucleation(Hammadi *et al.*, 2015) because it allows multiple independent experiments.

Manual handling of the sample crystals can mechanically and environmentally stress them. The stress induced during this delicate step may affect crystal quality and decrease its diffractive power. To minimize manual handling, an alternative is in situ X-ray data collection. One method involves using X-ray-transparent microfluidic devices (Hansen *et al.*, 2006, Dhouib *et al.*, 2009, Stojanoff *et al.*, 2011, Guha *et al.*, 2012, Pinker *et al.*, 2013, Khvostichenko *et al.*, 2014, Horstman *et al.*, 2015, Heymann *et al.*, 2015, Maeki *et al.*, 2015). Another solution following Yadav's pioneering work (Yadav *et al.*, 2005), is to collect X-ray data directly in micro capillary (Li *et al.*, 2006, Maeki *et al.*, 2012). For ex situ data collection, Gerdts (Gerdts *et al.*, 2010) and Stojanoff (Stojanoff *et al.*, 2011) harvested a protein crystal from a microfluidic chanel using a cryo-loop and Li(Li *et al.*, 2006) made crystals flow out of a capillary, then looped and flash-froze them.

We present an application that addresses these two problems using a microfluidic platform developed in our group.(Zhang *et al.*, 2017) We optimize crystallization conditions of human Quinose Reductase 2 (QR2 EC 1.10.5.1)(Nosjean *et al.*, 2000), and collect X-ray data both in situ and ex situ to characterize the crystals obtained.

## 2. Optimization and crystallization results using the microfluidic platform

The microfluidic platform developed in our group offers four modular functions(Zhang *et al.*, 2017): droplet formation, on-line UV characterization, incubation and observation (figure 1). We adapt the platform to generate droplets of 2nL in long Teflon tubing (150µm ID from IDEX Health and Science), without using surfactant.(Zhang *et al.*, 2015) Droplets are generated by crossing a continuous phase (FC70 fluorinated oil from Hampton research) with dispersed phases (containing the protein and the crystallization agent(s)) in a microfluidic junction (Te, cross or 7-entry junction from IDEX Health and Science). A programmable syringe pump (neMESYS, cetoni GmbH) controls the flow-rates of the different fluids. We couple an on-line UV detector (USB2000+, Ocean optics) to the Teflon tubing after the droplet formation zone ((3) in figure 1). We use on-line analysis of the droplets by UV spectrometry to characterize the chemical composition gradient generated among droplets of identical sizes.(Zhang *et al.*, 2017)

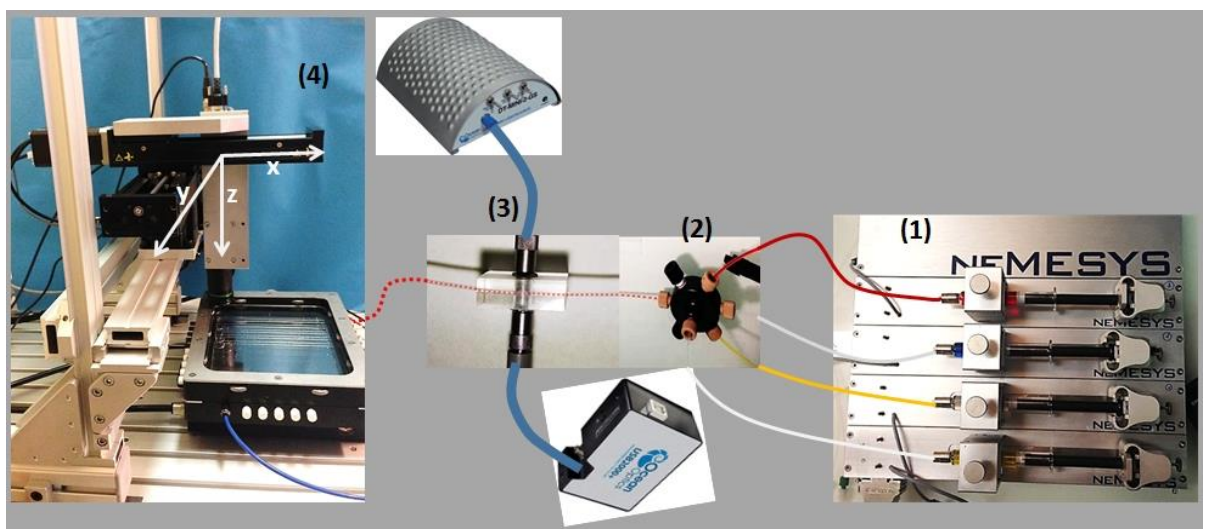

**Figure 1.** Pictures of the home-made microfluidic platform: (1) syringe pump, (2) 7-entry junction, (3) on-line UV module (4) tubing-holder for thermostatting and observation with XYZ-motorized camera.

Experimental conditions are based both on solubilities obtained by equilibrating crystal-solution suspensions over time (figure 1, Supplementary Information) and crystallization conditions used for structural determination.(Foster *et al.*, 1999) Subsequent gradient optimization, using experimental conditions presented in figure 2, provides optimal conditions leading to high quality crystals. At least 60 droplets of 2nL per experimental conditions were generated and observed (figure 2). Crystals in droplets from experiment (b) (figure 2) were used for X-ray diffraction (XRD).

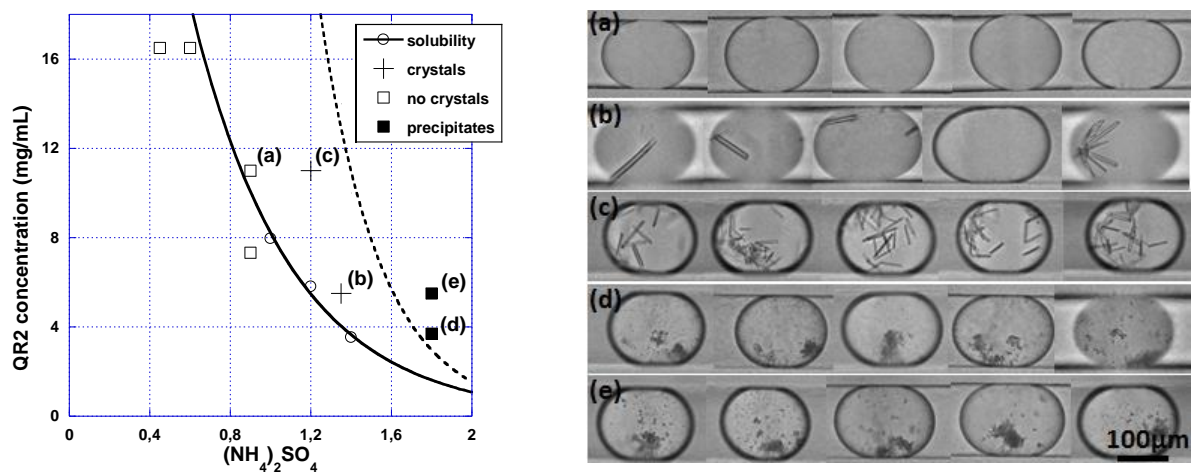

**Figure 2.** Solubility of QR2 versus $(NH_4)_2SO_4$ at pH=8 (20mM Tris-HCl and 150mM NaCl), and the different experimental conditions tested in the fine-gradient experiment. The dashed line is a guide for the eye to separate the crystallization and precipitation zones.
Photos of 5 representative droplets obtained in 2nL droplets in Teflon capillary (150µm ID) after 24h. (a) 11 mg/mL QR2, 0.9M $(NH_4)_2SO_4$, (b) 5.5 mg/mL QR2, 1.35M $(NH_4)_2SO_4$, (c) 11 mg/mL QR2, 1.2M $(NH_4)_2SO_4$, (d) 3.7 mg/mL QR2, 1.8M $(NH_4)_2SO_4$, (e) 5.5 mg/mL QR2, 1.8M $(NH_4)_2SO_4$, at 20°C.

## 3. XRD characterization

Although direct X-ray data collection from the microfluidic devices is used to minimize manual handling, Teflon-related background noise is significant on diffraction patterns. This may reduce the quality of the diffraction data (Yadav *et al.*, 2005) and strongly reduce the observed diffraction limits of the crystals. Hence, we tested two approaches: (1) transferring droplets containing the crystals of interest from Teflon to silica tubing for in situ XRD without freezing; and (2) extracting the crystals of interest from the tubing, depositing them on a MicroMesh$^{TM}$, a polyimide grid transparent to X-rays, for ex situ XRD thereby avoiding mechanical shocks.

**3.1. In situ XRD.** We transferred the droplets from experiment (b), performed in Teflon tubing, to silica tubing (fused silica tubing with a polyimide coating -150µm ID, 360µm OD, from IDEX health and science) using a linear junction (IDEX health and science). The internal silica tubing wall was coated with a commercial hydrophobic surface-coating agent (Aquapel®, PPG industry)(Mazutis *et al.*, 2009) to ensure droplet stability. The silica tubing containing the droplets was directly mounted on a magnetic base extracted from standard SPINE sample loops, ready for transfer to the X-ray setup (figure 3). For the proof of concept, a single crystal was analysed by XRD at room temperature on the beam line PROXIMA 1 (Synchrotron SOLEIL). Diffraction was observed to a resolution of 2.7 Å (figure 2, Supplementary Information). However, the strong X-ray damage to the crystal from these room-temperature measurements most likely explains why a complete data set was not obtained from one single crystal.

Microfluidics, however, can produce hundreds to thousands of droplets with identical composition. Thus, serial-crystallography at room temperature would yield a complete set of data for structural resolution with limited noticeable effects from radiation damage. This approach was used recently by Heymann(Heymann *et al.*, 2015) with a chip made of PDMS and COC (cyclic olefin polymer) or Kapton.

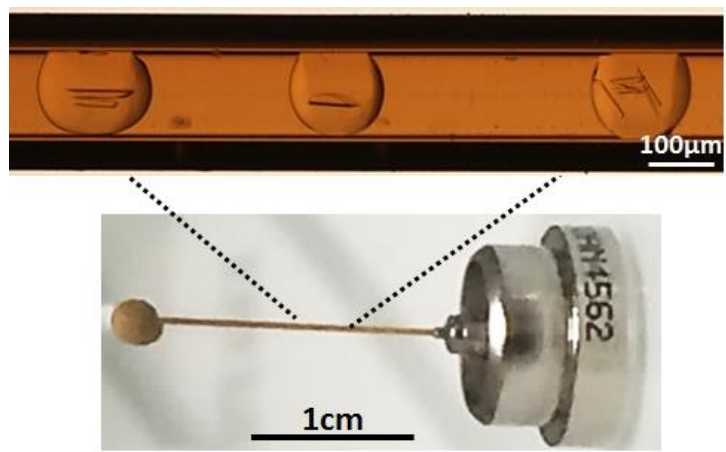

**Figure 3.** Photos of the silica tubing mounted on a magnetic base.

**3.2. Ex situ XRD.** Here, crystals were harvested from the Teflon tubing containing droplets. A droplet was deposited on a Micromesh$^{TM}$ (MiTeGen) using a high-precision micro-injector for flow control (Elveflow). The micro-injector and the MicroMesh$^{TM}$ are fixed to home-made micromanipulators for precise displacement in X, Y and Z(Grossier R. *et al.*, 2011) (figure 3, Supplementary Information). Crystals were placed singly on the MicroMesh$^{TM}$ (figure 4 and figure 4 and video 1, Supplementary Information) which was immediately extracted from the oil bath (FC70) and immersed in liquid nitrogen to cryogenize the crystals. Here, FC70 oil acted as a cryoprotectant, but crystals can be immersed in a drop of glycerol for cryoprotection. Then, XRD was carried out. By extracting the crystals without direct handling or mechanical stress and preparing the sample for diffraction studies under cryogenic conditions, we were able to collect a full data set at a resolution of 2.3Å (with or without glycerol). By determining structure from one single crystal, we identified electron density for the Flavin Adenine Dinucleotide (FAD) cofactor in the active site of QR2 (figure 5). Further studies should explore screening of QR2 co-crystallization with ligands for structure-based drug design. These first results confirm that the microfluidic approach yields crystallographic data of sufficient quality to allow us to judge whether or not the ligands bind to the active site.

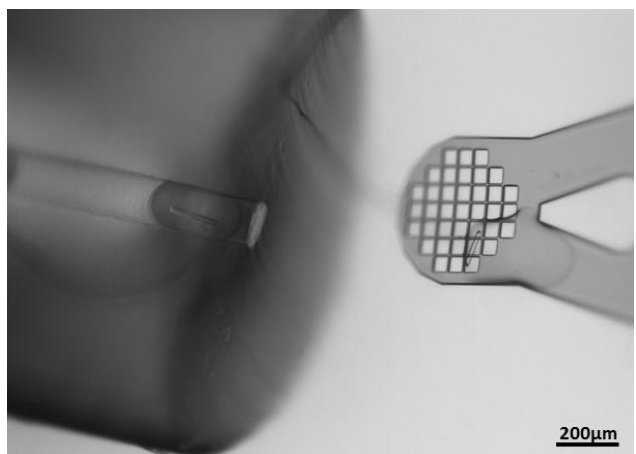

**Figure 4.** Photo of a crystal in a droplet deposited on the Micromesh$^{TM}$.

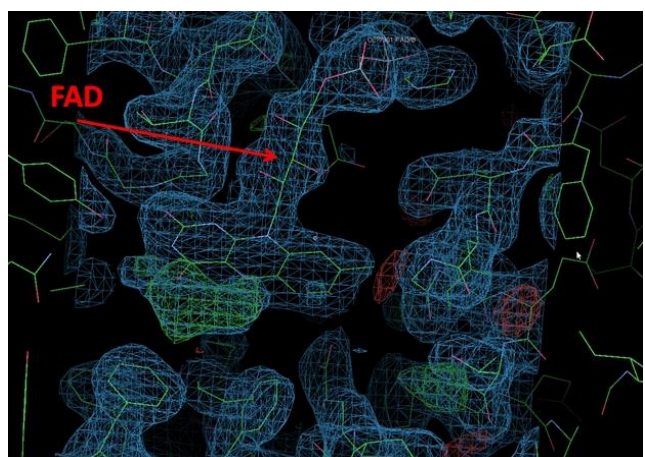

**Figure 5.** Electronic density map of the active site of QR2, with the FAD cofactor.

## 4. Conclusions

We present an application of a microfluidic platform developed in our group to the optimization of crystallization conditions of the pharmaceutical protein QR2. The resulting crystals were characterized by both in situ and ex situ X-ray diffraction.

**Acknowledgments**

We thank the Institut de Recherche Servier for financial support. We thank T. Bactivelane (CINaM), M. Lagaize (CINaM) and M. Audiffren (ANACRISMAT) for technical assistance. We thank G. Sulzenbacher, S. Spinelli and P. Cantau (AFMB) for the transport of the crystals and their support about crystallography. Experiments at Synchrotron SOLEIL were performed under the *in house* proposal number 99150097. Results incorporated in this note received funding from the European Union's Horizon 2020 research and innovation programme under grant agreement No 708130.